\documentstyle[prl,aps,floats,graphicx]{revtex}

\begin{document}
\preprint{hep-ph/0008113}
\draft
  
\twocolumn[\hsize\textwidth\columnwidth\hsize\csname 
@twocolumnfalse\endcsname

\title{ Primordial black hole production due to preheating}
\author{Anne M.~Green,$^1$ and Karim A.~Malik$^2$}
\address{$^1$Astronomy Unit, School of Mathematical
Sciences, Queen Mary and Westfield College, Mile End Road, London, E1
4NS, Great Britain}
\address{$^2$Relativity and Cosmology Group, 
School of Computer Science and Mathematics,
University of Portsmouth,\\Portsmouth PO1 2EG, Great Britain}
\date{\today} 
\maketitle
\begin{abstract}
During the preheating process at the end of inflation the
amplification of field fluctuations can lead to the amplification of
curvature perturbations.  If the curvature perturbations on small
scales are sufficiently large, primordial black holes (PBHs) will be
overproduced.  In this paper we study PBH production in the two-field
preheating model with quadratic inflaton potential.  We show that for
many values of the inflaton mass $m$, and coupling $g$, small scale
perturbations will be amplified sufficiently, before backreaction can
shut off preheating, so that PBHs will be overproduced during the subsequent
radiation dominated era.
\end{abstract}

\pacs{PACS numbers: 98.80.Cq \hfill hep-ph/0008113}

\vskip2pc]

\section{Introduction}
There has recently been great interest in the
reheating process which occurs at the end of inflation. It has been
found that in many inflation models reheating proceeds initially via a
period of broad parametric resonance, known as {\em
preheating}~\cite{genpre,KLS97}. During preheating the inflaton field decays
extremely rapidly producing an exponentially growing number of
particles, until the backreaction of the particle production shuts off
the parametric resonance with the particles produced subsequently
thermalising and the universe becoming radiation dominated.

Attention has been focused on two-field models with either a
quadratic~\cite{KLS97} or a quartic inflaton
potential\cite{GKLS,Kaiser}. It has been claimed in Ref.~\cite{Betal}
that the amplification of curvature perturbations could be large with
fluctuations being driven non-linear, even on large `super-Hubble'
scales. This would have serious consequences for cosmology, destroying
the standard picture of structure formation~\cite{LL}, in particular
by violating the microwave background constraint on the amplitude of
the power spectrum on large scales, and leading to the overproduction
of primordial black holes (PBHs) on small scales. The growth on large
scales, relevant for structure formation, has been confirmed for the
quartic case~\cite{Bruce+Viniegra,Fin+Brand,Zibin}, whilst it has
recently been shown for the quadratic inflaton
potential~\cite{jedam,ivan,LLMW}, that in fact the amplification of
the fluctuations on large scales is negligible. The formation of PBHs
has not been studied for either model. For the quadratic potential the
power spectrum of curvature perturbations has a $k^{3}$-spectrum
however, growing strongly towards small scales, so that PBH formation
may still be problematic despite the amplification of large scale
fluctuations being negligible.

In this letter we present the first study of the formation of PBHs due
to the resonant amplification of field fluctuations during preheating.
The scales which we are interested in pass outside the Hubble radius
towards the end of inflation.  The fluctuations are amplified during
preheating and PBHs will be formed, when these scales re-enter the
Hubble radius after reheating during the radiation dominated era, if
the fluctuations are sufficiently large.  PBHs may in principle also
be formed on the slightly shorter scales which re-enter the Hubble
radius during preheating, when the universe is dominated by the
oscillating scalar fields, however 3-D simulations along the lines of
the 1D simulations carried out in Ref.~\cite{EP} would be necessary to
study this.

\section{Calculation of the power spectrum}

In common with
Refs.~\cite{Betal,LLMW} we will study the two-field model with
quadratic scalar field potential
\begin{equation}
V(\phi,\chi) = \frac{1}{2} \, m^2\phi^2 +  \frac{1}{2} \, 
g^2\phi^2\chi^2 \,,
\end{equation}
where $\phi$ is the inflaton and $\chi$ the preheating field, $m$ the
mass of the inflaton and $g$ the coupling of the inflaton to the
preheat field. The $\phi$ field, and hence the effective mass of the
$\chi$ field, $m_{\chi}=g \phi$, oscillates around zero with large
amplitude resulting in efficient preheating. The power spectrum of the
curvature perturbation in the constant density gauge, $\zeta$,
produced in this model has recently been calculated by Liddle, Lyth,
Malik and Wands~\cite{LLMW} (LLMW hereafter). The scales which we are
interested in have $k \ll g \Phi$, where $\Phi$ is the initial amplitude of
the $\phi$-field oscillations, and are outside the Hubble radius
at the end of inflation, so that we can use the results of their
calculation.  In this section we outline the relevant results of LLMW,
for further details see Ref.~\cite{LLMW}.

Strong parametric resonance, resulting in large amplification of the
initial quantum fluctuation in the $\chi$ field, occurs if
\begin{equation}
q \equiv \frac{g^2 \Phi^2}{4m^2} \gg 1 \,.
\end{equation}

The power spectrum of a quantity  $x$ is defined as~\cite{LL}
\begin{equation}
{\cal P}_{x} \equiv \frac{k^3}{2 \pi^2} \left\langle|x_{{\bf k}}|^2
        \right\rangle \,,
\end{equation}
where $k = |{\bf k}|$ is the comoving wavenumber, $x_{{\bf k}}$ are
the coefficients of the Fourier expansion of $x$, and the average is
over ensembles.  The effect of preheating on the amplitude of the
$\chi$ field can be modelled as~\cite{KLS97,LLMW}
\begin{equation}
{\cal P}_{\delta \chi} =  \left. {\cal P} _{\delta \chi} 
              \right|_{{\rm end}} 
             \exp{ \left(2 \mu_{k} m \Delta t \right)} \,,
\end{equation}
where $\Delta t$ is the time elapsed since the end of inflation.
The power spectrum of $\delta\chi$ at the end of inflation 
is denoted by $\left. {\cal P}_{\delta \chi} \right|_{{\rm end}}$.
The Floquet index $\mu_k$ is taken as 
\begin{equation}
\label{defmuk}
\mu_k\simeq\frac{1}{2\pi} 
        \ln \left( 1+ 2e^{-\pi\kappa^2} \right) 
\,,
\end{equation}
and
\begin{equation}
\label{defkappa}
\kappa^2
\equiv \left(\frac{k}{k_{{\rm max}}}\right)^2
\equiv \frac{1}{18\sqrt{q}}\left(\frac{k}{k_{{\rm end}}}\right)^2 
\, ,
\end{equation}
where $k_{{\rm max}}$ is the scale which undergoes maximum
amplification and $k_{{\rm end}}$ is the comoving wavenumber of the scale
which exits the Hubble radius at the end of inflation. For strong
coupling ($q \gg 1$) $\mu_{k} \approx \mu = \ln{3}/2 \pi \approx 0.17$
for all modes outside the Hubble radius at the end of inflation
($\kappa \ll 1$).

LLMW find that the evolution of the curvature perturbations is due 
to the non-adiabatic pressure perturbation $\delta p_{\rm{nad}}$,
which is dominated by second-order fluctuations in $\chi$, 
$g^2 \phi^2 \delta \chi^2$ (see also~\cite{WMLL}).  
The first-order contribution, which is of order $g^2
\phi^2 \chi \delta \chi$, is negligible since the background field
$\chi$ is vanishingly small~\cite{jedam,ivan}.  The power 
spectrum of $\zeta_{\rm
nad}$ resulting from the amplification of the non-adiabatic
fluctuations in the $\chi$ field, is then given by 
\begin{eqnarray}
\label{spectrum}
{\cal P}_{\zeta_{\rm nad}} 
\simeq A
\left({k\over k_{\rm
end}}\right)^3 I(\kappa,m\Delta t) \,,
\end{eqnarray}
where
\begin{equation}
A={2^{9/2}3 \over \pi^5\mu^2} \left({\Phi \over m_{\rm Pl}}\right)^2
\left({H_{\rm end} \over m}\right)^4 g^4 q^{-1/4} \, ,
\end{equation}
and
\begin{equation}
\label{integral}
I(\kappa,m\Delta t)\equiv \frac{3}{2} \int_0^{\kappa_{\rm cut}} \!\! d\kappa'
\int_0^\pi \! d\theta \, e^{2(\mu_{\kappa'}+\mu_{\kappa-\kappa'})m\Delta t} 
\kappa'^2 \sin\theta \,,
\end{equation}
where $\theta$ is the angle between ${\bf k}$ and ${\bf k}'$ and 
$\kappa_{\rm cut}$ is an ultraviolet cut-off,  $\kappa_{\rm cut} 
\sim\kappa_{\rm max}$. 
LLMW found that on the scales
relevant for large scale structure formation the change in $\zeta$ due to
preheating is negligible, however since ${\cal P}_{\zeta_{\rm nad}}
\propto k^{3}$ smaller scale fluctuations with wavenumbers 
$k \sim k_{{\rm end}}$ may become large enough for PBHs to be 
overproduced, before backreaction shuts off preheating.

The duration of the first stage of preheating, $m\Delta t_{\rm{BR}}$,
during which backreaction is unimportant and fluctuations are
amplified exponentially, can be estimated numerically by finding the
smallest solution of~\cite{KLS97}
\begin{equation}
\label{predur}
m\Delta t \approx \frac{1}{4\mu}\ln\frac{10^{6} (m \Delta t)^3 }
               {g^4\sqrt{48\pi q}} \,.
\end{equation}
After this stage backreaction becomes
important; energy is drained from the inflaton field, and transferred
to the $\chi$ field, diminishing the amplitude of the inflaton
oscillations.   
The growth in the preheating field changes the effective mass of the $\chi$ 
field itself, thereby rendering preheating less efficient.
The growth in the $\chi$ field also changes the oscillation 
frequency of the inflaton, narrowing the resonance band and 
eventually shutting the resonance down. The duration of
this second stage of preheating is typically far shorter than that of
the first stage of preheating.  The solution of Eq.~(\ref{predur}),
$m\Delta t_{\rm{BR}}$, thus provides us with a conservative estimate of
the duration of the entire preheating process.

\section{Black hole abundance constraints}

Primordial black holes may be formed in the the early universe via the
collapse of sufficiently large density perturbations~\cite{carr}. Due
to their observational consequences there are tight constraints on
their abundance; typically less than $10^{-20}$ of the energy in the
universe can go into PBHs at the time that they form~\cite{pbh,gl}.
For a PBH to form during radiation domination the density contrast,
$\delta$, when the fluctuation enters the horizon must exceed a
critical value $\delta_{{\rm c}}$, where numerical simulations find
$\delta_{{\rm c}} \sim 0.7$~\cite{nj}. The fraction of the energy of
the universe which goes into PBHs is found by integrating the
probability distribution of the fluctuations, $p(\delta)$, over the
large, PBH forming, fluctuations,
\begin{equation}
\label{b1}
\beta= \frac{\rho_{{\rm pbh}}}{\rho_{{\rm tot}}} \approx
         \int_{\delta_{{\rm c}}}^{\infty} p(\delta) {\rm d} \delta \,,
\end{equation}
and is controlled by the mass variance, $\sigma(k)$, at horizon 
crossing, of the fluctuations:
\begin{equation}
\label{s1}
\sigma^2(k) = \int_{0}^{\infty} \delta^2 p(\delta) {\rm d} \delta \,.
\end{equation}

The density perturbation distribution is close to gaussian; strictly
speaking it is a chi-squared distribution with a large number of
degrees of freedom, resulting from the convolution of the square of
the gaussian distributed field fluctuations over a wide range of
scales.\footnote{This is significantly different from the case of single
field models where difficulties arise in calculating
$\beta$~\cite{ng}. In those models PBH production requires large field
fluctuations, which are non-linear and therefore may not be gaussian 
distributed. Furthermore there
is no definite prescription for relating the field and metric
fluctuations if the field fluctuations are large.
This is not the case in our model as the leading order metric perturbations
arise from terms which are second-order in the \emph{linear} field 
fluctuations.}
We find, using Eqs.~(\ref{b1}) and (\ref{s1}), that for a gaussian
probability distribution PBHs are overproduced, $\beta > 10^{-20}$, if
$\sigma(k)> \sigma_{{\rm thresh}}=0.08$. For comparison for a first
order chi-squared distribution the threshold would be $\sigma_{{\rm
thresh}}=0.03$. We use $\sigma_{{\rm thresh}}=0.08$ as a
conservative estimate of the threshold for PBH overproduction.
\begin{figure}
\begin{center}
\includegraphics[angle=270,width=0.45\textwidth]{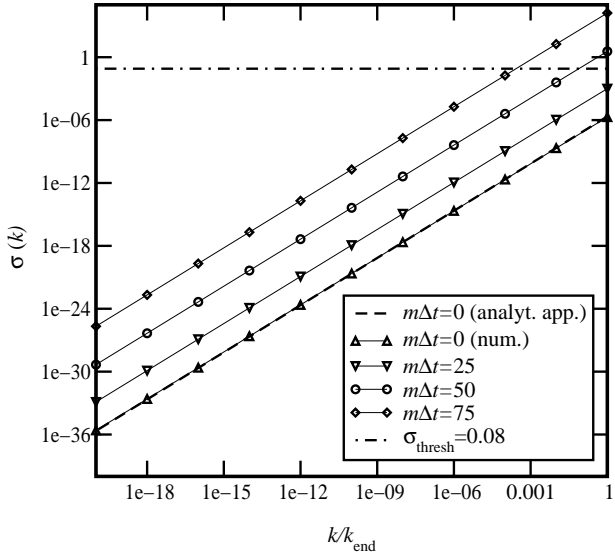}
\caption[fig1]{\label{fig1} The mass variance at horizon crossing,
$\sigma(k)$, versus dimensionless wavenumber $k/k_{\rm{end}}$, where
$k_{{\rm end}}$ is the comoving Hubble radius at the end of inflation,
for $m \Delta t=0,25, 50$ and $75$, for inflaton mass $m=10^{-6}$ and
coupling $g=10^{-3}$. The dot-dashed line shows the threshold
$\sigma_{\rm{thresh}}=0.08$ above which PBHs will be overproduced.}
\end{center}
\end{figure}
To calculate $\sigma(k)$ the power spectrum must be smoothed using a
window function, ${W}(kR)$,~\cite{LL}:
\begin{equation}
\label{defsigma}
\sigma^2(k) = \frac{16}{81} \int_{0}^{\infty} \left( 
        \frac{\tilde{k}}{k} \right)^4 {\cal P}_{\zeta_{{\rm nad}}}(\tilde{k})
        {W}^2(\tilde{k}R) \frac{{\rm d} \tilde{k}}{\tilde{k}} \,.
\end{equation}
The precise form of the window function is not important and we take it
to be a gaussian,
\begin{equation}
W(\tilde{k}R)=\exp{\left( - \frac{\tilde{k}^2 R^2}{2} \right)} \,,
\end{equation}
where $R = 1/k = 1/a H $, is the scale of interest, i.e. the Hubble
radius at the time the PBHs form.

For $m \Delta t=0$, i.e. immediately after inflation ends before
preheating commences, $I(\kappa,0)=\kappa_{\rm cut}^3\sim 1$, so 
that we can calculate $\sigma(k)$ approximately analytically. Inserting
Eq.~(\ref{spectrum}) into Eq.~(\ref{defsigma}) and defining the
dimensionless wavenumber $\tilde{\kappa}=\tilde{k}/k_{\rm{max}}$
we find
\begin{eqnarray}
\label{sigmat0}
\left.\sigma^2(k)\right|_{{m \Delta t=0}} \approx \frac{16}{81} A
\left(\frac{k_{\rm{max}}}{k}\right)^4
\left(\frac{k_{\rm{max}}}{k_{\rm{end}}}\right)^3  \times \\ \nonumber
\int^{\infty}_0 \tilde{\kappa}^6 \exp\left[-
\left(\frac{k_{\rm{max}}}{k} \right)^2  \tilde{\kappa}^2 
     \right] {\rm d} \tilde{\kappa} \, .
\end{eqnarray}
The integral in Eq.~(\ref{sigmat0}) above can be evaluated as
\begin{equation}
\int^{\infty}_0 \tilde\kappa^6 \exp\left[
\left(\frac{k_{\rm{max}}}{k} \right)^2 \tilde\kappa^2\right] 
d \tilde\kappa 
=\frac{15}{16}\sqrt{\pi}\left(\frac{k}{k_{\rm{max}}} 
\right)^7 \, ,
\end{equation}
so that the mass variance squared at the beginning
of preheating is given by
\begin{equation}
\label{scheck}
\left. \sigma^2(k) \right|_{{m \Delta t=0}} \approx 10\sqrt{2 \pi}A\,
         q^{\frac{3}{4}} \left(\frac{k}{k_{\rm{end}}} \right)^3 \,,
\end{equation}
which provides a check for our numerical results below.

\begin{figure}
\begin{center}
\includegraphics[angle=270,width=0.45\textwidth]{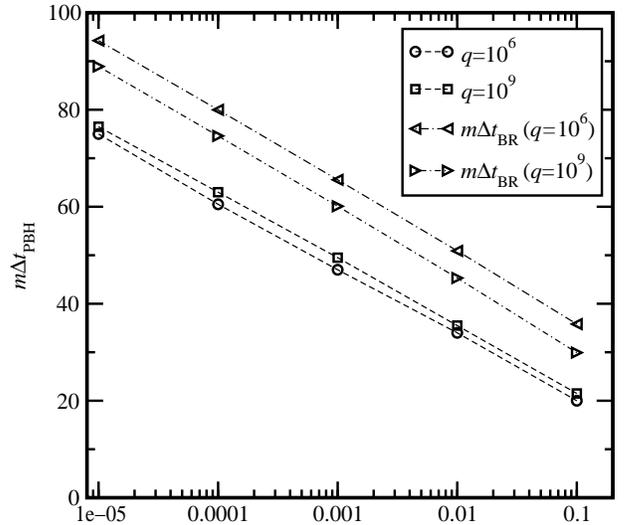}
\caption[fig2]{\label{fig2} The minimum preheating duration for which
PBHs are overproduced, $m\Delta t_{\rm{PBH}}$, versus the coupling $g$
for $q=10^6$ and $10^9$. The dot-dashed lines show the time at which
the first stage of preheating ends for the respective values of
$q$. This shows that the fluctuations are amplified sufficiently,
before backreaction sets in, for PBHs to  be
overproduced subsequently.}
\end{center}
\end{figure}

\section{Results}

We evaluate Eq.~(\ref{spectrum}) numerically using
the values of the Hubble parameter $H$ and the inflaton $\phi$ at the
end of slow roll inflation: $H_{\rm{end}}=m/3$ and
$\Phi=\phi_{\rm{end}}=m_{\rm{Pl}}/\sqrt{12\pi}$.

In Fig.~\ref{fig1} we plot the mass variance at horizon crossing,
$\sigma(k)$, for several values of $m \Delta t$, versus the
dimensionless wavenumber $k/k_{{\rm end}}$, for the commonly used
parameter values $m=10^{-6} m_{{\rm Pl}}$, $g=10^{-3}$; for these
values $q\approx 6600$.  For $m \Delta t=0$ we reproduce the analytic
expression found above, Eq.~(\ref{scheck}).  We find that for $m
\Delta t \sim 35$ the mass variance $\sigma(k)$ exceeds
$\sigma_{\rm{thresh}}=0.08$ on small scales, leading to the
overproduction of PBHs. Since $\sigma(k)$ is largest on small scales,
PBHs are formed most easily immediately after preheating. In this
preheating model $ k= a H \propto a^{-1/2}$ and the scale factor grows
by at most $\sim 16$~\cite{KLS97} so that $k$ changes by, at most, a
factor of 0.25 during preheating. For definiteness we therefore
evaluate $\sigma(k)$ on the scale $k_{\star}=0.25 k_{{\rm end}}$. This
approach is conservative; we over-estimate the Hubble radius
immediately after preheating and consequently also over-estimate the
preheating duration neccesary to overproduce PBHs.

\begin{figure}
\begin{center}
\includegraphics[angle=270,width=0.45\textwidth]{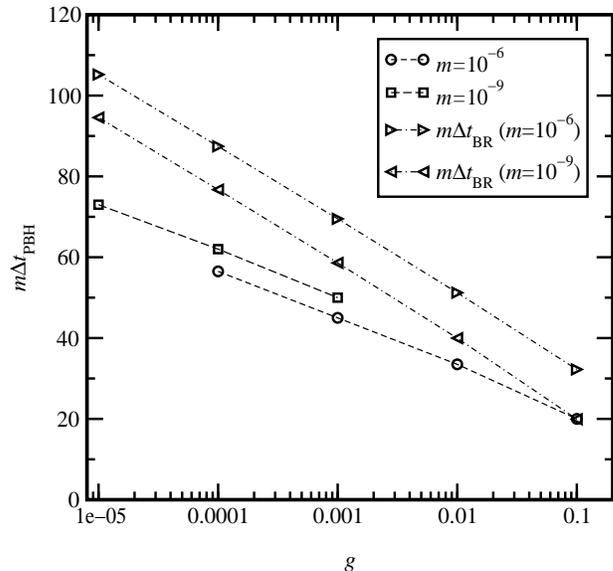}
\caption[fig3]{\label{fig3} The minimum preheating duration for which
PBHs are overproduced, $m\Delta t_{\rm{PBH}}$, versus the coupling $g$
for two values of the inflaton mass, $m=10^{-6}m_{{\rm Pl}}$ and
$10^{-9}m_{{\rm Pl}}$. The dot-dashed lines show the time at which
the first stage of preheating ends for the respective masses. This
shows that the fluctuations are amplified sufficiently, before
backreaction sets in, for PBHs to be overproduced subsequently.}
\end{center}
\end{figure}

In Fig.~\ref{fig2} we plot $m\Delta t_{\rm{PBH}}$, the minimum
preheating duration for which the mass variance on small scales
exceeds the threshold for PBH overproduction, $\sigma(k_{{\star}})>
\sigma_{{\rm thresh}}=0.08$, versus the coupling $g$, for two values
of the parameter $q$, $q=10^6$ and $10^9$. For comparison we also plot
the estimate of the duration of preheating, $m \Delta t_{{\rm BR}}$,
found from Eq.~(\ref{predur}). Similarly in Fig.~\ref{fig3} we plot $m
\Delta t_{\rm {PBH}}$ and $m \Delta t_{{\rm BR}}$ as a function of $g$
for $m=10^{-6} m_{{\rm Pl}}$ and $10^{-9} m_{{\rm Pl}}$. The minimum
preheating duration for which PBHs are overproduced, $m \Delta
t_{\rm{PBH}}$, depends most strongly on the strength of the coupling
between the inflaton and preheating fields, $g$, decreasing as the
coupling increases.  We find that, for all sets of parameters
investigated, the field fluctuations are amplified sufficiently during
preheating that PBHs are overproduced during the subsequent radiation
dominated era.  The minimum preheating duration for which PBHs are
overproduced is at least $m \Delta t\sim 25$ smaller than than the
duration of the first stage of preheating, during which backreaction
is unimportant and the fluctuations undergo exponential amplification.

\section{Discussion}

In this letter we have investigated primordial black hole formation in
the two-field preheating model with a quadratic inflaton potential.
Whilst the amplification of the field fluctuations on the scales relevant
for structure formation is negligible~\cite{jedam,ivan,LLMW}, the
power spectrum of curvature perturbations has a $k^{3}$-spectrum,
i.e. the fluctuations are largest on small scales.  We find that
the fluctuations are amplified sufficiently during preheating, 
before backreaction is expected to shut-off the resonance, for PBHs to 
be overproduced during the subsequent radiation dominated era. We 
emphasize that, in the strong 
preheating regime, the overproduction occurs irrespective of the strength 
of the coupling of the inflaton to the preheat field.
This constitutes a serious problem for this preheating model.

\section*{Acknowledgements.}
We thank Bruce Bassett, Karsten Jedamzik, Andrew Liddle and David Wands for 
useful discussions. 
AMG is supported by the PPARC and acknowledges use of the Starlink
computer system at QMW, KAM acknowledges the use of the Starlink
computer system at Sussex.

 
\end{document}